\input harvmac
\input graphicx
\input color

\def\Title#1#2{\rightline{#1}\ifx\answ\bigans\nopagenumbers\pageno0\vskip1in
\else\pageno1\vskip.8in\fi \centerline{\titlefont #2}\vskip .5in}

%
%
\ifx\includegraphics\UnDeFiNeD\message{(NO graphicx.tex, FIGURES WILL BE IGNORED)}
\def\figin#1{\vskip2in}
\else\message{(FIGURES WILL BE INCLUDED)}\def\figin#1{#1}
\fi
\def\Fig#1{Fig.~\the\figno\xdef#1{Fig.~\the\figno}\global\advance\figno
 by1}
%
%
%
%
\def\Ifig#1#2#3#4{
\goodbreak\midinsert
\figin{\centerline{
\includegraphics[width=#4truein]{#3}}}
\narrower\narrower\noindent{\footnotefont
{\bf #1:}  #2\par}
\endinsert
}

\font\ticp=cmcsc10

\def \purge#1 {\textcolor{magenta}{#1}}
\def \new#1 {\textcolor{blue}{#1}}
\def\comment#1{}

\def\\{\cr}
\def\text#1{{\rm #1}}
\def\frac#1#2{{#1\over#2}}

\def\hf{{1\over 2}} 
\def\calo{{\cal O}}
\def\cale{{\cal E}}

\def\eg{{\it e.g.}}
\def\roughly#1{\mathrel{\raise.3ex\hbox{$#1$\kern-.75em\lower1ex\hbox{$\sim$}}}}
\font\bbbi=msbm10 
\def\mathbb#1{\hbox{\bbbi #1}}

 \def\sch{Schr\"odinger}

\def\mthsu{\mathsurround=0pt  }
\def\leftrightarrowfill{$\mthsu \mathord\leftarrow\mkern-6mu\cleaders
  \hbox{$\mkern-2mu \mathord- \mkern-2mu$}\hfill
  \mkern-6mu\mathord\rightarrow$}
\def\overleftrightarrow#1{\vbox{\ialign{##\crcr\leftrightarrowfill\crcr\noalign{\kern-1pt\nointerlineskip}$\hfil\displaystyle{#1}\hfil$\crcr}}}
\overfullrule=0pt

%
%
\lref\Susstrouble{
  L.~Susskind,
  ``Trouble for remnants,''
[hep-th/9501106].
}
\lref\AMPS{
  A.~Almheiri, D.~Marolf, J.~Polchinski and J.~Sully,
  ``Black Holes: Complementarity or Firewalls?,''
  JHEP {\bf 1302}, 062 (2013).
  [arXiv:1207.3123 [hep-th]].
}
\lref\HaPr{
  P.~Hayden, J.~Preskill,
  ``Black holes as mirrors: Quantum information in random subsystems,''
JHEP {\bf 0709}, 120 (2007).
[arXiv:0708.4025 [hep-th]].
}
\lref\MaSu{
  J.~Maldacena and L.~Susskind,
  ``Cool horizons for entangled black holes,''
Fortsch.\ Phys.\  {\bf 61}, 781 (2013).
[arXiv:1306.0533 [hep-th]].
}
\lref\GiShtwo{
  S.~B.~Giddings and Y.~Shi,
  ``Effective field theory models for nonviolent information transfer from black holes,''
[arXiv:1310.5700 [hep-th]], Phys.\ Rev.\ D (in press).
}
\lref\NVNLT{
  S.~B.~Giddings,
  ``Modulated Hawking radiation and a nonviolent channel for information release,''
[arXiv:1401.5804 [hep-th]].
}
\lref\QBHB{
  S.~B.~Giddings,
  ``Quantization in black hole backgrounds,''
Phys.\ Rev.\ D {\bf 76}, 064027 (2007).
[hep-th/0703116 [HEP-TH]].
}
\lref\BHMR{
  S.~B.~Giddings,
  ``Black holes and massive remnants,''
Phys.\ Rev.\ D {\bf 46}, 1347 (1992).
[hep-th/9203059].
}
\lref\BHQIUE{
  S.~B.~Giddings,
  ``Black holes, quantum information, and unitary evolution,''
  Phys.\ Rev.\ D {\bf 85}, 124063 (2012).
[arXiv:1201.1037 [hep-th]].
}
\lref\SGmodels{
  S.~B.~Giddings,
   ``Models for unitary black hole disintegration,''  Phys.\ Rev.\ D {\bf 85}, 044038 (2012)
[arXiv:1108.2015 [hep-th]].
}
\lref\NLvC{
  S.~B.~Giddings,
  ``Nonlocality versus complementarity: A Conservative approach to the information problem,''
Class.\ Quant.\ Grav.\  {\bf 28}, 025002 (2011).
[arXiv:0911.3395 [hep-th]].
}
\lref\NLEFTone{
  S.~B.~Giddings,
  ``Nonviolent information transfer from black holes: a field theory parameterization,''
Phys.\ Rev.\ D {\bf 88}, 024018 (2013).
[arXiv:1302.2613 [hep-th]].
}
\lref\NVNL{
  S.~B.~Giddings,
  ``Nonviolent nonlocality,''
  Phys.\ Rev.\ D {\bf 88},  064023 (2013).
[arXiv:1211.7070 [hep-th]].
}
\lref\BHSM{
  S.~B.~Giddings,
  ``Statistical physics of black holes as quantum-mechanical systems,''
Phys.\ Rev.\ D {\bf 88}, 104013 (2013).
[arXiv:1308.3488 [hep-th]].
}
\lref\GiShone{
  S.~B.~Giddings and Y.~Shi,
  ``Quantum information transfer and models for black hole mechanics,''
Phys.\ Rev.\ D {\bf 87}, 064031 (2013).
[arXiv:1205.4732 [hep-th]].
}
\lref\WABHIP{
  S.~B.~Giddings,
  ``Why aren't black holes infinitely produced?,''
Phys.\ Rev.\ D {\bf 51}, 6860-6869 (1995).
[hep-th/9412159].
}
\lref\Hawk{
  S.~W.~Hawking,
  ``Particle Creation By Black Holes,''
  Commun.\ Math.\ Phys.\  {\bf 43}, 199 (1975)
  [Erratum-ibid.\  {\bf 46}, 206 (1976)].
}
\lref\LPSTU{
  D.~A.~Lowe, J.~Polchinski, L.~Susskind, L.~Thorlacius and J.~Uglum,
  ``Black hole complementarity versus locality,''
Phys.\ Rev.\ D {\bf 52}, 6997 (1995).
[hep-th/9506138].
}
\lref\UnWamine{ W.~G.~Unruh and R.~M.~Wald
  ``How to mine energy from a black hole,''
Gen.\ Relat.\ Gravit.\ {\bf 15}, 195 (1983).}
\lref\stringmine{
 A.~E.~Lawrence and E.~J.~Martinec,
  ``Black hole evaporation along macroscopic strings,''
Phys.\ Rev.\ D {\bf 50}, 2680 (1994)
[hep-th/9312127]\semi
  V.~P.~Frolov and D.~Fursaev,
  ``Mining energy from a black hole by strings,''
Phys.\ Rev.\ D {\bf 63}, 124010 (2001)
[hep-th/0012260]\semi
  V.~P.~Frolov,
 ``Cosmic strings and energy mining from black holes,''
Int.\ J.\ Mod.\ Phys.\ A {\bf 17}, 2673 (2002).
}
\lref\Locbdt{
  S.~B.~Giddings and M.~Lippert,
  ``The Information paradox and the locality bound,''
Phys.\ Rev.\ D {\bf 69}, 124019 (2004).
[hep-th/0402073].
}
\lref\BHV{
  S.~Bravyi, M.~B.~Hastings and F.~Verstraete,
  ``Lieb-Robinson Bounds and the Generation of Correlations and Topological Quantum Order,''
Phys.\ Rev.\ Lett.\  {\bf 97}, 050401 (2006).
}
\lref\LIGO{
  B.~P.~Abbott {\it et al.} [LIGO Scientific and Virgo Collaborations],
  ``Observation of Gravitational Waves from a Binary Black Hole Merger,''
Phys.\ Rev.\ Lett.\  {\bf 116}, no. 6, 061102 (2016).
[arXiv:1602.03837 [gr-qc]].
}
\lref\EHT{
  S.~Doeleman {\it et al.},
 ``Imaging an Event Horizon: submm-VLBI of a Super Massive Black Hole,''
[arXiv:0906.3899 [astro-ph.CO]].
}
\lref\Isit{
  S.~B.~Giddings,
  ``Is string theory a theory of quantum gravity?,''
Found.\ Phys.\  {\bf 43}, 115 (2013).
[arXiv:1105.6359 [hep-th]].
}
\lref\UQM{
  S.~B.~Giddings,
  ``Universal quantum mechanics,''
Phys.\ Rev.\ D {\bf 78}, 084004 (2008).
[arXiv:0711.0757 [quant-ph]].
}
\lref\BPS{
  T.~Banks, L.~Susskind and M.~E.~Peskin,
  ``Difficulties for the Evolution of Pure States Into Mixed States,''
Nucl.\ Phys.\ B {\bf 244}, 125 (1984)..
}
\lref\Haag{R. Haag, {\sl Local quantum physics, fields, particles, algebras}, Springer (Berlin, 1996).}
\lref\EREPR{
  J.~Maldacena and L.~Susskind,
  ``Cool horizons for entangled black holes,''
Fortsch.\ Phys.\  {\bf 61}, 781 (2013).
[arXiv:1306.0533 [hep-th]].
}
\lref\vanRent{
  M.~Van Raamsdonk,
  ``Building up spacetime with quantum entanglement,''
Gen.\ Rel.\ Grav.\  {\bf 42}, 2323 (2010), [Int.\ J.\ Mod.\ Phys.\ D {\bf 19}, 2429 (2010)].
[arXiv:1005.3035 [hep-th]].
}
\lref\ADM{
  R.~L.~Arnowitt, S.~Deser and C.~W.~Misner,
  ``Canonical variables for general relativity,''
Phys.\ Rev.\  {\bf 117}, 1595 (1960).
}
\lref\Isha{
  C.~J.~Isham,
 ``Canonical quantum gravity and the problem of time,''
[gr-qc/9210011].
}
\lref\AgAs{
  I.~Agullo and A.~Ashtekar,
  ``Unitarity and ultraviolet regularity in cosmology,''
Phys.\ Rev.\ D {\bf 91}, no. 12, 124010 (2015).
[arXiv:1503.03407 [gr-qc]].
}
\lref\Cortetal{
  A.~Corichi, J.~Cortez and H.~Quevedo,
  ``On the relation between Fock and Schrodinger representations for a scalar field,''
Annals Phys.\  {\bf 313}, 446 (2004)
[hep-th/0202070]\semi
J.~Cortez, G.~A.~Mena Marugán and J.~M.~Velhinho,
  ``Quantum unitary dynamics in cosmological spacetimes,''
Annals Phys.\  {\bf 363}, 36 (2015).
[arXiv:1509.06171 [gr-qc]].
}
\lref\SGpr{S.~B.~Giddings, work in progress.}
\lref\SGalg{
  S.~B.~Giddings,
 ``Hilbert space structure in quantum gravity: an algebraic perspective,''
JHEP {\bf 1512}, 099 (2015).
[arXiv:1503.08207 [hep-th]].
}
\lref\locbdi{
  S.~B.~Giddings and M.~Lippert,
  ``Precursors, black holes, and a locality bound,''
Phys.\ Rev.\ D {\bf 65}, 024006 (2002).
[hep-th/0103231].
}
\lref\DoGione{
  W.~Donnelly and S.~B.~Giddings,
  ``Diffeomorphism-invariant observables and their nonlocal algebra,''
Phys.\ Rev.\ D {\bf 93}, no. 2, 024030 (2016), Erratum: [Phys.\ Rev.\ D {\bf 94}, no. 2, 029903 (2016)].
[arXiv:1507.07921 [hep-th]].
}
\lref\DoGitwo{
  W.~Donnelly and S.~B.~Giddings,
  ``Observables, gravitational dressing, and obstructions to locality and subsystems,''
Phys.\ Rev.\ D {\bf 94}, no. 10, 104038 (2016).
[arXiv:1607.01025 [hep-th]].
}
\lref\GiWe{S.~B.~Giddings and S.~Weinberg, work in progress}
\lref\Hawksoft{
  S.~W.~Hawking,
  ``The Information Paradox for Black Holes,''
[arXiv:1509.01147 [hep-th]].
}
\lref\HPS{
  S.~W.~Hawking, M.~J.~Perry and A.~Strominger,
  ``Soft Hair on Black Holes,''
Phys.\ Rev.\ Lett.\  {\bf 116}, no. 23, 231301 (2016).
[arXiv:1601.00921 [hep-th]]\semi
``Superrotation Charge and Supertranslation Hair on Black Holes,''
[arXiv:1611.09175 [hep-th]].
}
\lref\GiSh{
  S.~B.~Giddings and Y.~Shi,
  ``Quantum information transfer and models for black hole mechanics,''
Phys.\ Rev.\ D {\bf 87}, no. 6, 064031 (2013).
[arXiv:1205.4732 [hep-th]].
}
\lref\Sussxfer{
  L.~Susskind,
  ``The Transfer of Entanglement: The Case for Firewalls,''
[arXiv:1210.2098 [hep-th]].
}
\lref\Hawkunc{
  S.~W.~Hawking,
  ``Breakdown of Predictability in Gravitational Collapse,''
Phys.\ Rev.\ D {\bf 14}, 2460 (1976).
}
\lref\Pageone{
  D.~N.~Page,
  ``Average entropy of a subsystem,''
Phys.\ Rev.\ Lett.\  {\bf 71}, 1291 (1993).
[gr-qc/9305007].
}
\lref\Pagetwo{
  D.~N.~Page,
  ``Information in black hole radiation,''
Phys.\ Rev.\ Lett.\  {\bf 71}, 3743 (1993).
[hep-th/9306083].
}
\lref\Brownmine{
  A.~R.~Brown,
  ``Tensile Strength and the Mining of Black Holes,''
Phys.\ Rev.\ Lett.\  {\bf 111}, no. 21, 211301 (2013).
[arXiv:1207.3342 [gr-qc]].
}
\lref\SGLIGO{
  S.~B.~Giddings,
  ``Gravitational wave tests of quantum modifications to black hole structure -- with post-GW150914 update,''
Class.\ Quant.\ Grav.\  {\bf 33}, no. 23, 235010 (2016).
[arXiv:1602.03622 [gr-qc]].
}
\lref\SGwind{
  S.~B.~Giddings,
  ``Possible observational windows for quantum effects from black holes,''
Phys.\ Rev.\ D {\bf 90}, no. 12, 124033 (2014).
[arXiv:1406.7001 [hep-th]].
}
\lref\Polcphone{
  J.~Polchinski,
  ``Weinberg's nonlinear quantum mechanics and the EPR paradox,''
Phys.\ Rev.\ Lett.\  {\bf 66}, 397 (1991)..
}
\lref\Pres{
  J.~Preskill,
  ``Do black holes destroy information?,''
  in proceedings of Black holes, membranes, wormholes and superstrings, Houston 1992
[hep-th/9209058].
}
\lref\GiNe{
  S.~B.~Giddings and W.~M.~Nelson,
  ``Quantum emission from two-dimensional black holes,''
Phys.\ Rev.\ D {\bf 46}, 2486 (1992).
[hep-th/9204072].
}
\lref\GiPs{
  S.~B.~Giddings and D.~Psaltis,
  ``Event Horizon Telescope Observations as Probes for Quantum Structure of Astrophysical Black Holes,''
[arXiv:1606.07814 [astro-ph.HE]].
}
\lref\SuTh{
  L.~Susskind and L.~Thorlacius,
 ``Gedanken experiments involving black holes,''
Phys.\ Rev.\ D {\bf 49}, 966 (1994).
[hep-th/9308100].
}
\lref\SGobs{
  S.~B.~Giddings,
  ``Observational strong gravity and quantum black hole structure,''
Int.\ J.\ Mod.\ Phys.\ D {\bf 25}, no. 12, 1644014 (2016).
[arXiv:1605.05341 [gr-qc]].
}
\lref\PsJo{
  D.~Psaltis and T.~Johannsen,
  ``Sgr A*: The Optimal Testbed of Strong-Field Gravity,''
J.\ Phys.\ Conf.\ Ser.\  {\bf 283}, 012030 (2011).
[arXiv:1012.1602 [astro-ph.HE]].
}
\lref\Mathur{
  S.~D.~Mathur,
  ``The Information paradox: A Pedagogical introduction,''
Class.\ Quant.\ Grav.\  {\bf 26}, 224001 (2009).
[arXiv:0909.1038 [hep-th]].
}
\lref\SGfrank{S.~B.~Giddings, presentations at the 2015 {\sl Karl Schwarzschild Meeting}, Frankfurt Institute for Advanced Studies, https://indico.cern.ch/event/382362/contributions/904861/
attachments/1133170/1618390/Giddings\_2015-Schwarzschild.pdf , 
https://indico.cern.ch
/event/382362/contributions/904896/attachments/1132185/1618395/ \break
Discussion\_slides\_of\_Giddings.pdf\ .}
\lref\Psvid{Quantum Structure of Astrophysical Black Holes, http://xtreme.as.arizona.edu/
$\sim$dpsaltis/?page\_id=2757\ .}
\Title{
\vbox{\baselineskip12pt  
}}
{\vbox{\centerline{Nonviolent unitarization:  basic postulates to } \centerline{soft quantum structure of black holes
}}}

\centerline{{\ticp 
Steven B. Giddings\footnote{$^\ast$}{Email address: giddings@ucsb.edu}
} }
\centerline{\sl Department of Physics}
\centerline{\sl University of California}
\centerline{\sl Santa Barbara, CA 93106}
\vskip.10in
\centerline{\bf Abstract}
A first-principles approach to the unitarity problem for black holes is systematically explored, based on the postulates of 1) quantum mechanics 2) the ability to approximately locally divide quantum gravitational systems into subsystems 3) correspondence with quantum field theory predictions for appropriate observers and (optionally) 4) universality of new gravitational effects.  Unitarity requires interactions between the internal state of a black hole and its surroundings that have not been identified in the field theory description; correspondence with field theory indicates that these are soft.  A conjectured information-theoretic result for information transfer between subsystems, partly motivated by a perturbative argument, then constrains the minimum coupling size of these interactions of the quantum atmosphere of a black hole.  While large couplings are potentially astronomically observable, given this conjecture one finds that the new couplings can  be exponentially small in the black hole entropy, yet achieve the information transfer rate needed for unitarization, due to the large number of black hole internal states.  This provides a new possible alternative to arguments for large effects near the horizon.  If universality is assumed, these couplings can be described as small, soft, state-dependent fluctuations of the metric near the black hole. Open questions include that of the more fundamental basis for such an effective picture.

\Date{}

\newsec{Introduction}

Forty years of the study of quantum properties of black holes\refs{\Hawk} has made it evident that they cannot be consistently described within local quantum field theory, and thus that their quantum description must be given in some different framework.
Apparently, this is not simply a short distance problem, since as viewed from a conventional spacetime description, avoiding breakdown of quantum mechanics seems to require information transfer or other modifications to physics on scales comparable to a black hole's size.  Such information transfer would violate the locality of local quantum field theory (LQFT). Some time ago it was proposed that we seriously consider such {\it macroscopic} violations of conventional locality as the way to save quantum mechanics\refs{\BHMR,\Locbdt}, and many researchers now appear to  concur that a semiclassical description of spacetime must receive some such modification at horizon scales, or greater.

An exciting development is that, almost concurrently with the broadened acceptance of this viewpoint, we have entered into a new observational era, where we now have {\it two}  ways to observationally probe physics near a black hole horizon: via gravitational waves, with LIGO/VIRGO\refs{\LIGO}, and via interferometry using Earth-sized baselines, with the Event Horizon Telescope\refs{\EHT}.   This suggests the possibility of observation providing guidance, through discovery of or constraints on new effects on these scales\SGwind.

The problem of unitarity of black hole evolution seems to represent a foundational crisis in present-day physics:  it exhibits a fundamental clash between the principles underlying LQFT.  These are the principles of quantum mechanics, the principles of relativity, and the principle of locality, and apparently one or more of these requires modification.  This raises the challenging question of how to make progress.  This paper -- as with earlier related work -- will take the approach of beginning with very general principles, and asking how, within those principles, this ``unitarity crisis" can be resolved.  This approach will be based on the assumption that quantum mechanics is valid, but will for example allow weakening of the LQFT principle of locality.  Nonetheless, we will assume that for many purposes LQFT gives a good {\it approximate} description of physics, and so such modifications to it can be parameterized in an ``effective field theory" approach.  

This paper will take an agnostic approach to the grand hope that string theory provides a complete theory of quantum gravity.  This hope first emerged from string theory's success at addressing nonrenormalizability -- a {\it short-distance} issue.  Our more modern understanding appears to say that the 
more fundamental issue for gravity is the {\it long-distance}  problem of unitarity -- which is generically probed in the high-energy regime of the theory.  Many have believed this issue will be resolved via the AdS/CFT correspondence, but questions surrounding how AdS/CFT accurately describes bulk physics\refs{\Isit} have lingered for nearly two decades, and attempts to address black hole evolution via AdS/CFT have led to seemingly paradoxical conclusions such as the existence of ``firewalls\refs{\AMPS}."  This paper is agnostic on the role of strings in that it represents an approach to describe physically correct black hole behavior, whether or not such a description ultimately arises from the AdS/CFT framework.

In order to take such a basic, ``first principles" approach, we begin with a physically-motivated set of assumptions.  These will be followed to their logical conclusions -- quantum modification of the space-time structure near a black hole.  These 
Postulates, on which the current picture is based, are very simple:

{\it Postulate I, Quantum mechanics:}  Physics respects the essential principles of quantum mechanics.\foot{In the gravitational context, these must be suitably generalized to, {\it e.g.}, remove fundamental reliance on a basic notion of time; see \refs{\UQM} for further discussion.}  These include the assumption that configurations are described by a linear space of states, $\cal H$, with an inner product, and the assumption that dynamics is unitary, at least in the sense of being described by a unitary S-matrix when working with states with appropriate asymptotic boundary behavior.

{\it Postulate II, Subsystems:}  The Universe can be divided into distinct quantum subsystems, and in particular into black hole and environment subsystems, at least to a good approximation.  In the latter division, there are ``interior states" describing the configuration of the black hole subsystem.

{\it Postulate III, Correspondence with LQFT:}   Observations made by small freely falling observers in weak curvature regimes are approximately well described by a local quantum field theory lagrangian.  This includes observations made by observers freely falling through the classical horizon radius $R$, on scales small as compared to $R$.
These and other near-inertial observers find a minimal departure from LQFT.

{\it Postulate IV, Universality:}  Departures from the usual LQFT description influence matter and gauge fields in a universal fashion.

Postulate I needs little explanation: given the difficulties of modifying quantum mechanics (see {\it e.g.} \refs{\BPS,\Polcphone}), we assume a quantum-mechanical framework for physics.  Postulate II is a weakened version of the usual locality of LQFT; in LQFT one has  a precise division into subsystems based on commuting subalgebras of observables associated with spacelike-separated regions\refs{\Haag}.  

Postulate III is the statement that LQFT is, for many purposes, a valid description, even for observers falling into a black hole (BH). It is a postulate that we should take a conservative approach, and look for a minimal modification to our best current framework for describing nature.  Other work now advocates rather extreme departures from LQFT, such as a wall of Planck-energy particles, and planckian curvature, at the horizon\refs{\AMPS} or a drastic modification of quantum mechanics and spacetime structure in which entanglement at much larger scales than $R$ is associated with spacetime connectedness\refs{\vanRent,\EREPR}.  Postulate III guides us to looking for the minimal modification required to be consistent with quantum mechanics.

Postulate IV will be explained further below.  It is motivated by the need to address gedanken experiments involving BH mining\refs{\UnWamine,\stringmine}, and by desire to find approximate consistency with BH thermodynamics with the Bekenstein-Hawking entropy formula, as will be discussed.

Taken together, Postulates III and IV can be though of as a sort of ``weak quantum equivalence principle."  Namely, they indicate that experiments in small, freely falling laboratories do not experience large effects on short time scales, and that the effects that they do experience couple universally.

Further explanation of these Postulates and some of their motivation will be described below.  The Postulates will be followed to their logical implications, which include significant modifications to dynamics in the vicinity of a BH, and specifically new interactions with the quantum atmosphere of a BH, or a kind of ``soft quantum structure" on BHs.  An important question, as we will describe, is whether this structure is strong enough to have observational consequences which could be found by our two new approaches  to probing the near-horizon, strong-gravity region of a BH:  gravitational wave detection, and very long baseline interferometry.  Thus, observation potentially can furnish further information about this structure.

If the resulting picture is not correct, this would appear to indicate that one of the Postulates is not valid.  That -- and the nature of the modification to the Postulates -- would in itself be very interesting.

It is  important to understand another aspect of the approach taken by this paper.  While the paper is based on the simple Postulates above, we don't presently know a complete physical theory respecting these Postulates.  Instead -- as with the historical development of quantum mechanics -- the approach being taken here is to use the constraints of presumed correct physical behavior to work towards such a fundamental theory, which in the end may look rather different than our current framework of LQFT (or string theory).  In the absence of the complete fundamental theory, we proceed by adopting an {\it effective} description, in which we parameterize departures from the closest thing we have to an established fundamental theory, LQFT.\foot{Here we follow and elaborate on an approach developed in \refs{\SGmodels\BHQIUE\GiShone\NVNL\NLEFTone\GiShtwo\NVNLT-\GiPs}.} Ultimately, if we have identified the correct physical behavior through the Postulates, this should provide important guidance towards the final, more fundamental, description of quantum spacetime and gravity.

In outline, this paper begins by giving a \sch\ picture description of BH evolution in the  LQFT framework.  Such a description -- based on a choice of time slices -- sets up a connection to familiar discussions in quantum information theory of key concepts such as transfer of quantum information between subsystems.  Section three then investigates the necessary modifications of the interactions between a BH subsystem and its environment, in order to respect unitarity, following Postulate I.  These include new couplings between the BH and  the fields external to the BH, that are required in order to transfer information.  Then, Postulates III and IV imply important restrictions on these couplings, that they are localized near the BH, soft, and universal, resulting in the soft quantum structure for a BH.

Section four then investigates constraints connected with formulation of a general problem in quantum information theory, namely that of how fast information transfers between two subsystems given specific couplings between them.  
An answer to this question will constrain the size of the couplings between BH and surroundings.  Unitarity establishes a benchmark rate for information transfer from the BH.  At first sight, this suggests couplings of unit strength, for example in the universal case effectively behaving as $\calo(1)$ metric fluctuations.  However, analysis based on a proposed answer to this problem suggests that the benchmark unitarization rate can be achieved via small couplings, due to the contribution of the very large number of BH internal states, and thus suggests a scenario where unitarity is achieved without major impact on matter in the BH vicinity.  Since such soft structure can be present at distances $\sim R$ outside the horizon, this question -- strong vs. weak fluctuations -- also may be investigated by observation, either by LIGO, or by very long baseline interferometry with the Event Horizon Telescope.  Section five closes with a brief discussion of such prospects, as well as of problems related both to the connection between unitarization and strength of fluctuations, and to the question of the more fundamental underlying physics.

\newsec{Schrodinger evolution of LQFT in a black hole background}

Our problem is to describe localization and transfer of information in BH evolution.  While much discussion of BH evolution uses Heisenberg picture quantities, a clearer match to other discussions of quantum information and its transfer can be made if we instead work in \sch\ picture.  Here we can consider a wavefunction describing the state of the combined system of BH and its environment, and investigate information transfer between them.  

\subsec{Evolution}

To make the use of \sch\ picture clearer, we begin by establishing control of BH evolution in this picture in the context of LQFT dynamics in a fixed BH background.  This is our best current description of BH evolution, though our present unitarity crises demonstrates that ultimately it is incomplete.  This description can however be taken as a starting point for modifications to current principles -- following the above Postulates -- that are needed to restore unitary BH evolution.  Specifically, in the next section these modifications will be parameterized as departures from the LQFT \sch\ evolution.

 \Ifig{\Fig\Slices}{An  Eddington-Finkelstein diagram of a BH, with a family of slices, labeled by $T$, that smoothly cross the horizon.  The case shown is that of ``nice" slices, which asymptote to an internal radius $R_n$; one may alternately consider ``natural" slices which intersect the strong-curvature region near $r=0$.}{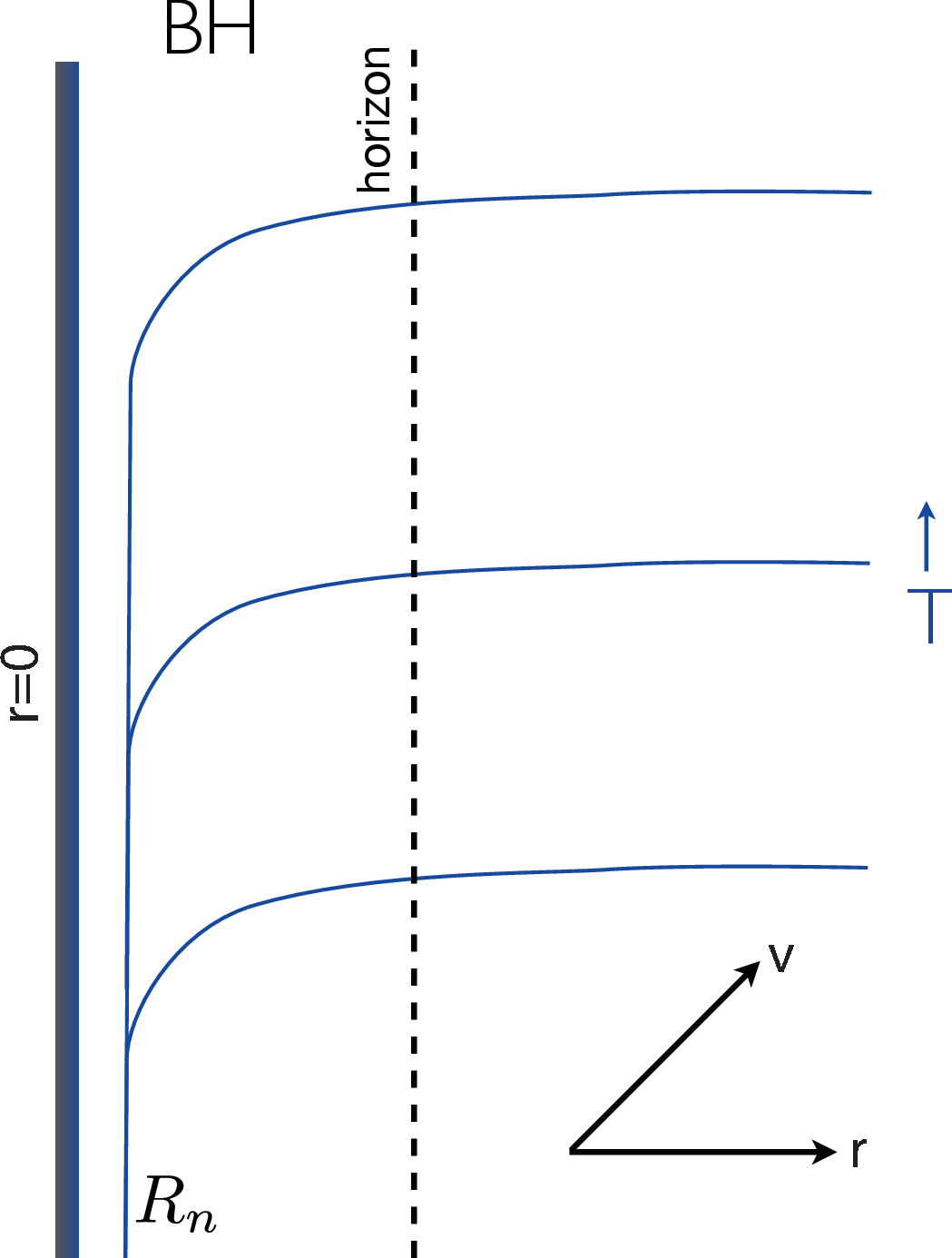}{3.5}

In order to describe Schr\"odinger evolution of the fields on a $D$-dimensional BH background, one needs to specify a time-slicing of the spacetime. Let a time parameter $T$ label the slices, and $x^i$ be coordinates along the slices.  We will be particularly interested in slices that smoothly cross the horizon and extend into the BH interior, as shown in \Slices.  Explicit examples of such slices are described in the Appendix.  In such a slicing, the metric takes the ADM form\refs{\ADM}
\eqn\ADMmet{ds^2= -N^2 dT^2 + q_{ij}(dx^i+N^i dT)(dx^j+N^j dT)\ ,}
where $N(T,x)$ and $N^i(T,x)$ are the lapse and shift, and $q_{ij}(x,T)$ is the spatial metric on the slices.

If we then consider, for example, a massless scalar field, with lagrangian
\eqn\sclagr{{\cal L} = - \hf g^{\mu\nu} \partial_\mu \phi \partial_\nu \phi}
the \sch\ evolution arises from the evolution operator
\eqn\schevo{U=\exp\left\{ -i \int dT H(T)\right\}}
where the hamiltonian is 
\eqn\slham{H(T) = \int d^{D-1} x \sqrt{q} \left[ \hf N(\pi^2 + q^{ij}\partial_i \phi \partial_j \phi )+ N^i \pi\partial_i\phi \right]\ .}
Here, $\pi$ is the canonical momentum conjugate to $\phi$.  This is defined as 
\eqn\canmom{\pi = {\partial_T \phi - N^i \partial_i\phi\over N}= n^\mu\partial_\mu\phi\ ,}
where 
\eqn\norma{n^\mu=(1,-N^i)/N}
is the unit normal to the slices,
and satisfies the commutators
\eqn\ccr{[\pi(x),\phi(x')] = -i{\delta^{D-1}(x-x')\over \sqrt{q}}\quad \leftrightarrow \quad \pi= -i {\delta \over \delta\phi}\ .}

It should be noted that in general curved spacetimes there can be subtleties with the use of \sch\ picture, as discussed, \eg, in \refs{\Isha\AgAs-\Cortetal}.  However, at the present we consider a black hole that is approximately static, since the geometry changes extremely slowly even with the emission of Hawking radiation.  Then, one can choose slicings (see the Appendix) where the metric is time independent.  This avoids these subtleties\refs{\SGpr}, which arise for time-dependent situations.

Beginning with a suitable initial state, the \sch\ evolution operator \schevo\ generates subsequent evolution, including the Hawking production\refs{\SGpr}, thus giving a systematic description of the LQFT dynamics.\foot{In particular, this offers an approach to deriving Hawking radiation that can be made less UV-sensitive than derivations following the original approach\Hawk.}  This, then, is a prototype for describing the more complete evolution of the BH.  

\subsec{Subsystems}

A next question is how Postulate II, the division into subsystems, is implemented.  For locally-finite quantum systems, subsystems are described by tensor factorization of the Hilbert space.  For field theory, however, such factorization is not possible due to the type-III property of the von Neumann algebras that arise there; colloquially, there is infinite local entanglement present in field theory states (for some further discussion and references, see \refs{\SGalg}).   So, instead, subsystems are defined by focusing on commuting subalgebras of the field algebra (see, \eg, \refs{\Haag}).  An example of such a subalgebra is that generated by field operators smeared with test functions with compact support in some open set in spacetime; such subalgebras, associated with spacelike-separated regions, commute.  One can, heuristically, think of subalgebras associated with different regions of spacetime as defining tensor factors of the Hilbert space, but this description is not fundamentally justified in continuum field theory.

Gravity presents new subtleties\refs{\locbdi,\Locbdt,\SGalg,\DoGione,\DoGitwo}.  Specifically, in the weak-gravity limit, the gauge symmetries of gravity are approximated by diffeomorphisms, which do not leave local operators invariant.  Gauge-invariant operators must be ``dressed;" colloquially, an operator creating a particle must also create its gravitational field.  This gravitational dressing moreover must generically extend to infinity\DoGitwo.  If one begins with commuting nongravitational operators in spacelike separated regions, and includes this dressing, then in general the operators no longer commute\refs{\locbdi,\Locbdt,\SGalg,\DoGione}.  This impedes extending the algebraic LQFT definition of subsystems to the gravitational setting.

While the usual algebraic definition of subsystems appears to fail in the gravitational context, for many purposes the failure appears to be small.  For example, if we consider two operators creating particles of energy $E$, at separation $r$, the non-vanishing commutator is characterized by the dimensionless quantity\DoGione
\eqn\commsize{{GE\over r^{D-3}}\ ,}
given by the locality bound\refs{\locbdi,\Locbdt}.  For an operator creating a particle inside a macroscopic BH together with an outside probe, one might therefore na\"\i vely expect a tiny effect, although details of this are being explored\refs{\GiWe}.  (If there {\it were} such an effect that plays an important role, by modifying the subsystem structure, it might be related to the ``soft hair" story advocated by \refs{\Hawksoft,\HPS}.)

For the purposes of this discussion, we will thus take a pragmatic approach, assuming that there is an approximate decomposition into subsystems
as in LQFT, and then will parameterize departure from the LQFT evolution.  While a more accurate description, including gravitational dressing, is ultimately needed, it may also be that the interactions parameterized in this paper can be used to capture essential implications of the nonlocality due to dressing.  

If we suppose we adopt such an approximate, pragmatic approach, we can straightforwardly divide the black hole and its environs into subsystems, \eg\ by considering operators, at a given time, localized inside and outside the horizon at $r=R$.  Thus, in this approximation, we have an example of how Postulate II is implemented.  

Alternatively, let us consider a slightly different division into subsystems, that does not place an artificial boundary at the horizon.  Specifically, consider a division into subsystems  with boundary between them at a more general radius $R_i$.  This could be less than $R$, for example  at $R_i=R/10$.  Thus,  for a given $T$, the subsystems correspond to the regions $r>R_i$ and $r<R_i$. 
Equal-$T$ operators associated with these regions will commute (neglecting dressing effects), and correspondingly we can heuristically think of the regions inside and outside $R_i$ as associated with different tensor factors of the full Hilbert space.

\subsec{Interactions between subsystems}

The hamiltonian \slham\ can likewise be divided up into pieces corresponding to these regions, so that the evolution can be described as unitary evolution of two coupled subsystems.  Specifically, 
\eqn\hamdiv{H= H_< + H_> + H_i\ ,}
where 
\eqn\hamdefs{\eqalign{H_<&=\int_{r<R_i} d^{D-1} x \sqrt{q} \left[ \hf N(\pi^2 + q^{ij}\partial_i \phi \partial_j \phi )+ N^i \pi\partial_i\phi \right]\cr  H_>&=\int_{r>R_i} d^{D-1} x \sqrt{q} \left[ \hf N(\pi^2 + q^{ij}\partial_i \phi \partial_j \phi )+ N^i \pi\partial_i\phi \right]\ ,}}
 and $H_i$ is an interaction term, at $R_i$, between the two subsystems.  Note that for $R_i<R$, due to field-theory causality, $H_>$ only transfers information {\it into} the boundary at $R_i$, and $H_<$ only transfers information from this boundary {\it further inward}.\foot{Such transfer of information can be characterized in terms of transfer of entanglement\refs{\HaPr\GiSh-\Sussxfer}.}  There is no transfer of information from inner subsystem to outer, since in LQFT information propagates within the light cones.

This is a perfectly consistent description of evolution of quantum fields on the black hole background.  It may be extended to other fields, \eg\ of gauge theory, or even corresponding to gravitational perturbations.  In principle one needs to deal with the singularity at $r=0$.  One possible approach to this is to introduce a set of states residing there ``into which information falls," but another alternative is simply to arrange the slices to never intersect $r=0$.  This can be done if all slices asymptote to an inner radius $R_n<R_i$, as in the nice-slice construction of \refs{\BHQIUE}.  Then, evolution freezes the fields at this radius\refs{\QBHB}, due to vanishing of the lapse $N$.  Effectively, this introduces a Hilbert space at $r=R_n$ in which infalling information accumulates.

\newsec{Interactions with the quantum atmosphere: soft quantum structure on BHs}

The LQFT evolution we have just described is  consistent and complete in the context of evolution on spatial slices on a fixed background BH geometry.  In particular, the evolution operator \schevo\ will be unitary from slice to slice.  However, since the evolution yields Hawking radiation, shrinking the black hole, the approximation of fixed background ultimately fails.  Since the black hole ultimately evaporates and disappears,\foot{Here, we assume no remnants, in accord with the arguments of \refs{\Pres\WABHIP-\Susstrouble}.} the preceding description can no longer be unitary in that context.  Specifically, if there is no information transfer from the interior subsystem to the environment of the BH, and if the BH ultimately disappears, unitarity has failed\refs{\Hawkunc}.  Since this contradicts Postulate I,  modification to the preceding description must be found.  This leads to the question: what are the minimal modifications to the standard LQFT description needed to restore unitarity?  The search for such a minimal departure is guided by the other Postulates.

\subsec{Unitarization through modified evolution}

As in the preceding section, we assume that Postulate II holds, so that the BH and its environment correspond to subsystems of the larger system, to a good approximation.  Again, we work slightly more generally, taking this division to occur at a radius $R_i$ which can be interior to the black hole.  Then the BH subsystem includes those excitations sufficiently deep in the BH.  This avoids artificially singling out the horizon.

If the BH subsystem disappears at the end of BH decay, any information stored in this subsystem must be emitted as this subsystem shrinks, in order to respect the unitarity of Postulate I.  As noted above, we can characterize transfer of information in terms of transfer of 
entanglement\refs{\HaPr\GiSh-\Sussxfer}.  Specifically, Hawking's calculations shows that the internal states of the BH are entangled with ``early" outgoing radiation that has been emitted; for an explicit two-dimensional example see \GiNe.  For the ultimate evolution to be unitary, this entanglement must transfer into later outgoing degrees of freedom, since no entanglement can remain with the BH once it has evaporated away.  Arguments by Page\refs{\Pageone,\Pagetwo} in particular tell us that when the BH has reached approximately the midpoint of its evaporation, its von Neumann entropy must stop increasing and begin to decrease.  In order to decrease to zero, this must be at a rate corresponding to approximately one qubit emitted per light-crossing time $R$.  

This emission of information (or transfer of entanglement) does not occur in the LQFT description of the preceding section, as was noted there.  Thus, Postulates I and II imply that there {\it must} be couplings between the subsystems that are not given by the previous LQFT description.  Moreover, the BH subsystem in the LQFT description is infinite dimensional, in conflict with this picture where its size gradually decreases to zero.  Together, these indicate that, at a minimum, we should modify the LQFT description so that 1) there are couplings between the BH and environment which can transfer information outward and 2) the BH subsystem is modeled as a finite-dimensional subsystem.

So, taking the approach of minimally modifying the dynamics, and respecting our Postulates, we parameterize departures from LQFT evolution by assuming that the BH subsystem is finite dimensional, the operator $H_<$ of \hamdefs\ gets replaced by some consistent evolution operator acting on this subsystem, and the interaction term $H_i$ gets replaced by a new interaction hamiltonian that yields transfer of information from BH states to those of the environment.  

Note that in order to respect Postulate III, that an infalling observer sees minimal departure from the predictions of LQFT, the subsystem at $r>R_i$ should be approximately well-described by LQFT degrees of freedom, and by the LQFT evolution given by $H_>$ of \hamdefs.  Moreover, $H_i$ should not introduce modifications to this which are too strong. 

Specifically, consider a BH that has evaporated sufficiently long that it is returning information (described as above in terms of entanglement) to its environment -- {\it e.g.} it is significantly past its half-life.  Of course, such a BH can still be very large, and its Hawking decay rate  still tiny; if it has a radius $R$, it emits energy $\sim 1/R$ in a time $R$, or, equivalently, the fractional change in mass during this time is $1/RM\sim 1/S_{BH}$, where $S_{BH}$ is the Bekenstein-Hawking entropy.  This appears to justify a treatment by a time-independent hamiltonian, to an excellent approximation, avoiding  subtleties noted in section II.  Since  the BH is in the phase of decreasing von Neumann entropy, it must also be emitting information at a rate of order one qubit per time $R$, so that its accumulated information $\sim S_{BH}$  shrinks to zero.  In this sense, $H_i$ must introduce an $\calo(1)$ correction to the Hawking emission -- for each Hawking particle emitted, there must also be $\calo(1)$ qubits of information emitted.  An important question is how to reconcile this with Postulate III, the approximate validity of LQFT. 

\subsec{Constraints from Postulate III}

In order to investigate properties of $H_i$, we first give a more careful description of the Hilbert space.  Working at such a ``late" time $T$, we can label states of the combined BH plus environment subsystems  as $|K,M;\psi_e,T\rangle$.  Here $M$ is the present mass of the BH, and $K$ ranges over the finitely-many states of the BH interior in a range of energies $\Delta E$ about $M$.  For generality, we parameterize the number of states in a range $\Delta E=1/R$ as $\exp\{S_{bh}\}$; it is widely believed that $S_{bh}\approx S_{BH}$.  The label $\psi_e$ parameterizes the state of the environment system, which, at least in the LQFT approximation, is thought of as residing at $r>R_i$.  This part of the state is approximately well-described as a state of quantum field theory, {\it e.g.} as described in the preceding section.

Since the internal part of the Hilbert space is now finite dimensional, with dimension $N\sim \exp\{S_{bh}\}$, the internal hamiltonian $H_<$ should now be a generator of $U(N)$, replacing the LQFT expression $H_<$ in \hamdefs.  For the moment, we will not need to make further assumptions about its structure, although of course it should approximately match onto LQFT evolution in relevant limits/regimes.  

The  term $H_i$ in the hamiltonian \hamdiv\ describes interaction between the internal BH states and the BH surroundings.  In LQFT, this term was localized at $R_i$; then, since information at $r<R$ only propagates inward, it only transferred information from $r>R_i$ to $r<R_i$.  In order to restore unitarity, given the ultimate disappearance of the BH, the more complete $H_i$  must also transfer information from the internal states to the region $r>R$, so that this information can escape the BH.  We thus write
\eqn\hisum{H_i= H_{i<}+ H_I\ ,}
where $H_{i<}$ is an analog of the original $H_i$ of LQFT, that only transfers excitations near $R_i$ into the internal Hilbert space, and $H_I$ is a new contribution to the hamiltonian, needed to transfer information outward.

Such interactions that transfer information out couple operators $\lambda^A$ that generate general $U(N)$ transformations on the interior states to LQFT operators with some support at $r>R$.  In order to parameterize the minimal departure from LQFT of Postulate III, we assume that the couplings are to sums of local operators; for example, products of local operators would introduce more significant departures, in the form of nonlocality in the surroundings of the black hole.  So, we parameterize $H_I$ as
\eqn\Hinew{H_I=\sum_{Ab}\lambda^A \int d^{D-1}x \sqrt{q}\, G_{Ab}(x) O^b(x)}
where $O^b(x)$ are local operators of the QFT,  $G_{Ab}(x)$ are coefficient functions that depend on the BH background, and integration is over $r>R_i$

Key questions in determining the effects of the new couplings \Hinew\ are the behavior of the coefficient functions $G_{Ab}(x)$, and that of which operators they couple to.  Answers are constrained by the Postulates.

First, if we seek a minimal deviation from LQFT, following Postulate III, we expect that the support of the $G_{Ab}(x)$ should be near the BH.  Eq.~\Hinew\ represents a departure from LQFT which ``nonlocally" transfers information, and clearly this is a more extreme departure if it extends far from the BH.  Let $R_a$ be the characteristic radius to which $G_{Ab}(x)$ are nonvanishing; unless other conditions dictate differently, Postulate III then implies $R_a\sim R$.  

But, too much localization gives violent deviation from LQFT.  For example, if the $G_{Ab}(x)$ vanish outside a microscopic distance $\epsilon$ outside the horizon, they must vary on this microscopic distance scale.  This implies couplings to the LQFT operators near the BH that inject hard momentum $\sim 1/\epsilon$, as seen by infalling observers.  For infalling observers not to see such ``hard" violations of LQFT near a large BH, we therefore require that the variation instead occurs over a scale $L$ that vanishes as $R\rightarrow\infty$.  The simplest possibility is that $L\sim R_a-R \sim R$, but one could have more general dependence, {\it e.g.} $R_a-R\sim R^p$, $L\sim R^q$, for some $p\geq q>0$.  Note that, if we do not wish to introduce a new scale besides that given by the BH size, the former, simplest, choice should be made; this also could be understood as imposing the condition of minimal violation of LQFT.  Similar comments apply to the angular variation of the $G_{Ab}(x)$.  Terms with larger angular momentum introduce harder deviations from LQFT, indicating that we should seek the minimum angular momenta necessary.  

The constraints of Postulate III also apply to the time dependence, which arises from the interaction between the interior hamiltonian $H_<$ and the $\lambda^A$  that nontrivially couple.  Specifically, if we go to a basis that diagonalizes $H_<$, the $\lambda^A$ will induce transitions among these energy eigenstates.  In order that $H_I$ not produce high-energy quanta in the vicinity of the BH, it should induce transitions with limited energy difference.  Again, the simplest possibility is that the $G_{Ab}$ describe transitions in energy $\delta E \roughly<1/R$.  This matches the energy scale of the Hawking radiation, and larger energies would represent a more extreme departure from LQFT.  The time dependence of \Hinew\ can  be directly seen by converting from \sch\ picture to an interaction picture, via the evolution operator for $H_<$. 

These combined conditions on the energy and momenta transitions arising from $G_{Ab}(x)$ are the conditions that correspond to the
 ``nonviolence" of the title.

\subsec{Universality}

We next turn to the motivation, meaning, and implications of Postulate IV, Universality.  In principle, $H_I$ could incorporate couplings to any local operators in the vicinity of the BH, and for example, linear couplings to a scalar field $\phi(x)$ have been considered as a toy model in \refs{\GiShtwo}.  However, some simple arguments imply powerful constraints on these couplings.

The first is the beautiful story of BH thermodynamics, with the BH entropy given, at least approximately, by the Bekenstein-Hawking entropy $S_{BH}$.  If we assume this story is preserved,
 this implies that  one can bring a BH into thermal equilibrium with fields at the Hawking temperature, $T_H$.  Generic couplings \Hinew\ will disturb this picture\refs{\BHSM}.  One way to think about this is via detailed balance.  In LQFT, the equilibrium is attained by balance between the inward flow of energy arising from the ``inward" $H_i$ of the preceding section, and the outward flow of energy of the Hawking radiation, found from $H_>$.  If the new $H_I$ contained operators that, for example, coupled to a specific field, then there would be additional outward flux of that field alone, violating the equilibrium condition.  

The second is the story of BH mining\refs{\UnWamine,\stringmine}.  The BH decay rate  can be increased by introducing non-trivial field configurations near the BH.  While Unruh and Wald\UnWamine\ consider rather complicated configurations, the simplest possibility is to ``thread" a cosmic string through the BH\refs{\stringmine}.  This introduces an additional channel for energy to escape, via modes on the string.  Moreover, it has been argued that many of the more general constraints from mining can be summarized within the simple story of cosmic string mining\refs{\Brownmine}.  The constraint on our hamiltonian arises because an increase in the rate of energy emission by the BH must be accompanied by an increase in the rate of transfer of information out of the BH -- otherwise, mining can induce a situation where a BH disappears before the full amount of its information is transferred out\refs{\NVNL}.  If the energy emission rate is increased by new modes that can be radiated (excitations on the string), a natural way to match these rates is if these modes can also transfer information from the BH\refs{\NLEFTone}.  This means that $H_I$ should couple to all such cosmic string modes, which is most simply achieved if it universally couples to all fields\NVNLT.

One naturally couples universally to all fields through the stress tensor,\foot{This is also expected to contain a contribution from gravitons.} motivating the restriction of \Hinew\ to interactions of the form
\eqn\HiT{H_I = \sum_A \lambda^A \int d^{D-1} x \sqrt{q} G_A^{\mu\nu}(x) T_{\mu\nu}(x) \ ,}
described now by the couplings $G_A^{\mu\nu}(x)$.  Recall that we have separated off an ``inward" piece of $H_i$ in \hisum.
The hamiltonian $H_I$ can alternately be written more concisely in the form
\eqn\Hiop{H_I=\int d^{D-1} x \sqrt{q} H^{\mu\nu}(x)T_{\mu\nu}(x)\ ,}
where $H^{\mu\nu}(x)$ is an operator acting on the internal BH Hilbert space.

It should be emphasized that such couplings do not {\it necessarily} satisfy the detailed 
balance necessary for equilibration, with entropy 
$S_{bh}=S_{BH}$.  For example, they could create outgoing excitations with energies above the thermal mean $\sim T_H$.  Assuming such equilibrium characterized by $S_{bh}\approx S_{BH}$ (at least to within $\calo(1)$ factors) thus further constrains these couplings, and in fact apparently reinforces arguments for the minimal choice described above, that the typical energy scales of the couplings are $\sim 1/R$.  

It also seems satisfying if the resolution to the problem of information loss has a universality, as seen in \HiT, that mirrors the known universality of gravity.  
The presence of \HiT, \Hiop\ indicates a form of ``soft" quantum gravitational structure of black holes, which we will explore further.

\newsec{Information transfer requirements}

To summarize the preceding discussion, we have considered evolution in a \sch\ picture.  States at time $T$ (in a definite slicing, or gauge, labeled by time at infinity) are assumed to be of the form $|K,M;\psi_e,T\rangle$.  Here we assume (Postulate II) an approximate subsystem division, where the labels $K$ describe the state of the ``internal" subsystem of the BH, which has dimension $N$, and the 
 labels $\psi_e$ describe the state of an environment subsystem, which is approximately well-described by LQFT degrees of freedom, and which may more generally extend a limited distance into the BH interior.  Evolution of these states is given by a hamiltonian of the form $H=H_<+H_>+H_i$, where $H_<$ generates certain $U(N)$ transformations, $H_>$ is LQFT evolution on the degrees of freedom described by $\psi_e$ -- as in \hamdefs\ -- and $H_i$ couples the two subsystems. $H_i$ in particular has a term coupling infalling excitations to the interior subsystem, as in LQFT (see \hisum), but also, as required by Postulate I, couplings  of the form \Hinew\ which can transfer information from  interior to environment.  Postulate III further restricts these couplings:  they have support near the BH, and involve soft energy/momentum scales, vanishing as $R\rightarrow\infty$.  The simplest assumption is that these scales are $\calo(1/R)$.  If Postulate IV is adopted, it then indicates couplings just to the energy-momentum tensor, as in \HiT.   One of its motivations, BH thermodynamics, also reinforces the case for energy scales of size $1/R$.  

Postulate I enforces further requirements on the couplings \HiT, since these couplings must transfer information {\it sufficiently rapidly}
 to ensure that all the information is transferred from the interior subsystem by the time the BH evaporates.  Formulating and applying this constraint leads us to an example of a more general problem in quantum information theory, which we turn to next.

\subsec{A problem in quantum information theory:  information transfer between subsystems}

Suppose that we have two quantum subsystems, $A$ and $B$, with Hilbert spaces of large dimensions, $|A|\gg1$, $|B|\gg1$.  We also assume $|B|\gg|A|$, though this could be generalized.  These subsystems are assumed to evolve by a time-independent hamiltonian of the form
\eqn\hcoup{H= H_A + H_B + H_I\ ,}
where $H_A$ acts on $A$, $H_B$ on $B$, and $H_I$ couples the two.  We can specifically assume that $H_I$ is of the form
\eqn\hidef{H_I = \cale \sum_{\gamma=1}^\chi c_\gamma O_A^\gamma O_B^\gamma\ .}
Here $\cale$ sets a common energy scale, and $c_\gamma$ are dimensionless coefficients.  
$O_A^\gamma$ are independent ({\it e.g.} commuting) unit-norm operators acting on $A$, and likewise for $O_B^\gamma$ on $B$.  
To define the norms, assuming random matrices, we use the standard operator norm
\eqn\Onorm{\Vert O_A^\gamma\Vert =  {\rm sup}  \vert O_A^\gamma\Psi\vert\ ;}
with the maximum taken over states $\Psi$ subject to the constraint $\vert \Psi\vert=1$.
For large  $|A|$ and random $O_A^\gamma$, this scales the same with  $|A|$
as the norm 
\eqn\OnormF{\Vert O_A^\gamma\Vert_s^2= {1\over |A|} {\rm Tr}[( O_A^{\gamma})^2]\ }
(note the latter differs from the standard Frobenius norm by the factor $1/|A|$).

Now, suppose we begin in a state where $A$ is thought of as containing a large amount of information; {\it e.g.} $A$ might be in a state entangled with another subsystem $\bar A$.  This can be maximized if $\bar A$ is a copy of $A$ that is maximally entangled with $A$.  Then, the question is how  the rate of transfer of information from $A$ to $B$ (which can be defined via rate of transfer of the entanglement with $\bar A$) depends on $H_A$, $H_B$, and $H_I$, as well as other aspects of the states?

This general problem seems not to have been fully addressed in the literature.\foot{Though, related bounds were discussed in \BHV.} But, with a few further assumptions, relevant to our problem, a conjecture can be formulated characterizing the information transfer rate.  It is of interest to further sharpen such a statement.

First, let us assume that $H_A$ and $H_B$ are generic generators of $U(|A|)$ and $U(|B|)$, for example
\eqn\HAexp{H_A= {\cal E} \sum_a h_a \lambda^a}
with the same energy 
$\cal E$ as above, setting the scale,  where $h_a$ are some general real dimensionless coefficients, {\it e.g.} with $\sum_a (h_a)^2/|A|=1$, and $\lambda^a$ is a basis of generators of $U(|A|)$.  Similar statements could be made for $B$, with a basis $\lambda^\beta$.

We would like to know how the entanglement transfer rate depends on the 
parameters of \HAexp\ and \hidef.  
This can be defined in terms of the rate of transfer of mutual information of $B$ and $\bar A$,
\eqn\mutinf{I({\bar A}:B)= S_{\bar A}+S_B-S_{{\bar A} B}\ ,}
where entropies are defined as the von Neumann entropy of the respective density matrices, formed from partial traces.\foot{One can alternately work with the mutual information $I({\bar A}:A)$; for further discussion see \GiShone.}
At early times, before the systems equilibrate, one expects linear growth with time.  A conjecture is that this rate behaves as
\eqn\entrate{{dI\over dt} = C {\cal E} \sum_{\gamma=1}^\chi c_\gamma^2\ ,}
with some dimensionless constant $C$, and for small enough $c_\gamma$.\foot{I thank W. van Dam and C. Nayak for discussions on the question of sharpening this conjecture.} 

To understand this conjecture, note first that, by energy conservation, $\cale$ sets the scale of the energy transfer from $A$ to $B$, and sets a nominal scale for the transfer rate.  One can then think of the quantities $c_\gamma$ as couplings for $\chi$ different information transfer ``channels" labelled  by $\gamma$.  If such a coupling gives the amplitude for a transfer interaction, the rate is proportional to its square; additional perturbative discussion appears in the next section.  Proving -- or improving -- the conjecture \entrate\ is an interesting problem in information theory, for future work.  We will use the conjecture as a starting point for discussing constraints on the couplings in the interaction hamiltonian \Hinew\ or \HiT.  

\subsec{Information transfer from black holes}

We next apply the preceding arguments to our hamiltonian \hamdiv, with modified $H_i$ now given by \hisum.  The information transferring $H_I$ of \Hinew\ is of the form of eq.~\hidef. 
The nonviolence condition from Postulate III will then restrict the allowed functions $G_{Ab}(x)$ to those that have low momentum and connect states with small energy differences, {\it e.g.} with both scales set by $\sim 1/R$.    This has the effect of   limiting the number of channels that contribute.  Then, the conjecture \entrate, together with the information transfer rate needed for unitarity, $\sim 1/R$, sets a minimum size for the couplings, for a given set of operators.

We explore this constraint within the context of the universal couplings of Postulate IV, although the discussion can be extended to  more general couplings like \Hinew.  The coupling functions $G_A^{\mu\nu}(x)$ of \HiT\ are directly seen from that equation to be dimensionless.  Comparing with  \hidef, we take the characteristic energy scale to be $\cale\sim 1/R$.  Nonviolence implies that there is a limited set of functions $G_A^{\mu\nu}(x)$ that play a role.  Let $f^{\mu\nu}_\gamma(x)$ be a basis of such sufficiently low-momentum (soft) functions, indexed by $\gamma=1,\ldots,\chi$.  Then expand
\eqn\Gexp{G_A^{\mu\nu}(x)=\sum_\gamma c_{A\gamma} f^{\mu\nu}_\gamma(x)\ ,}
with dimensionless expansion coefficients $c_{A\gamma}$.  Eq.~\HiT\  then takes the form
\eqn\HIexp{H_I= \sum_{\gamma A} \lambda^A c_{A\gamma} \int d^{D-1} x \sqrt{q} f_\gamma^{\mu\nu}(x) T_{\mu\nu}(x)\ .}
Comparing with \hidef\ shows that the operators
\eqn\intops{O_\gamma = \sum_A \lambda^A c_{A\gamma}}
are analogous to the $c_\gamma O^\gamma_A$ in \hidef, and that the operators 
\eqn\extops{ T_\gamma=\int d^{D-1} x \sqrt{q} f_\gamma^{\mu\nu}(x) T_{\mu\nu}(x)}
are analogous to  $\cale O^\gamma_B$ in \hidef.

We can now apply the expression \entrate\ to estimate the information transfer rate.  
Some caution is needed since the $T_\gamma$ are in general unbounded operators.  However, beginning with a given external state of the BH ({\it e.g.} a standard BH vacuum) on which $T_\gamma$ can act, we can consider a set of states that are in a range of energies $\cale$ around this state.  Then, a norm \Onorm\ can be found on this {\it subspace} of the full space of states. The subspace restriction eliminates the very high-energy states; these should not be relevant, due to energy conservation, as will be seen momentarily in a perturbative discussion.  We then choose the 
$f_\gamma^{\mu\nu}$ so that the operators $T_\gamma/\cale$ are unit norm.  Then we expect an equation of the form \entrate\ to hold, giving
\eqn\entrateBH{{dI\over dt} = {C \cale} \sum_{\gamma } \Vert O_\gamma\Vert^2 \ .}
For $\cale\sim 1/R$, this will be the necessary entanglement transfer rate
\eqn\necrate{{dI\over dt}\sim {1\over R} }
if
\eqn\opnorm{\sum_{\gamma=1}^\chi \Vert O_\gamma\Vert^2 \sim 1\ .}

To understand the latter condition, suppose that we can regard the operators $\calo_\gamma$ as behaving like random $N\times N$ matrices; after all, we expect the internal BH evolution to be rather chaotic, and there is no obvious reason to expect the couplings in \intops\ to be simple in an eigenbasis of energy $H_<$.  For random matrices, as previously noted, the operator norm scales with $N$ the same as the square norm \OnormF.  So, if $\lambda^A$ are normalized so that Tr$(\lambda^A\lambda^B)=\delta^{AB}$, the normalization \opnorm\ corresponds to the condition
\eqn\coupnorm{\sum_{A\gamma} c^2_{A\gamma} \sim N\ ,}
or, to couplings of size $c_{A\gamma}\sim \sqrt{1/N\chi}$.  

This scaling, and thus the conjecture \entrate, can be motivated by a perturbation theory argument.  Suppose that the BH is taken to initially have an internal state $|\psi\rangle$ which behaves like a random state ({\it e.g.} random superposition of eigenstates of $H_<$).  Then, we can estimate the rate for a transition from $|\psi\rangle$ to another state with energy difference $\sim 1/R$, during which the BH emits a quantum created by $T_\gamma$ of \extops\ acting on the exterior state, via Fermi's Golden Rule.  This rate takes the form
\eqn\decrate{\Gamma \approx {2\pi} \omega^{bh}(E) \sum_\gamma |\langle K|O_\gamma|\psi\rangle|^2\ |\langle\beta| {T_\gamma}|\alpha\rangle|^2\ ,}
where $E\approx M-1/R$ is the final state BH energy, $\omega^{bh}(E)$ is the BH density of states, $\langle K|O_\gamma|\psi\rangle$ is a matrix element with a typical final BH state $|K\rangle$ of energy $E$, and $|\alpha\rangle$ and $|\beta\rangle$ are the initial and final states of the exterior.   The density of states behaves as $\omega^{bh}(E)\sim NR$.   With the normalizations \opnorm, \coupnorm, the matrix element $\langle K|O_\gamma|\psi\rangle$ for a random $O_\gamma$ has typical size $1/\sqrt {N\chi}$, translating into an $\calo(1)$ transition rate in time $R$.  With the expectation that for each such transition $\calo(1)$ qubit of information is emitted, this gives the needed rate \necrate.

While these arguments have been given for the stress-tensor couplings \HiT, both for simplicity and because of the motivations for Postulate IV, they can clearly be extended to more general couplings of the form \Hinew.  If such more general couplings were relevant, this would likewise relate their required strengths to the rate \necrate\ needed for unitarization.

Either with couplings via $T_{\mu\nu}$, or via more general operators, note the origin of the unitarization rate \necrate.  The matrix elements of the operators  $O_\gamma$ coupling to the BH internal states can be tiny, $\sim 1/\sqrt N\sim \exp\{-S_{bh}/2\}$.  But, the rate $dI/dt$ can nonetheless be $\calo(1)$, due to the enormous factor $N$ in the density of states for the BH, enhancing the total transition rate.  Small couplings are effectively amplified by the enormous number of BH states.

\subsec{Size of effective metric fluctuations}

Universal couplings like \HiT\ or \Hiop\ can be interpreted as quantum contributions to an effective metric, which is a perturbation of the BH metric.  Specifically, \Hiop\ suggests that the metric is effectively perturbed by
\eqn\Metpert{\Delta g_{\mu\nu} = 2 H_{\mu\nu}(x)\ ,}
where indices are lowered with the background BH metric.  However,
recall that  $H^{\mu\nu}$ is operator-valued, so the effect of these 
interactions is not quite as simple as a classical shift in the metric, and in particular $H^{\mu\nu}$ depends on the state of the BH.  An important question is how this modifies evolution of matter in the BH vicinity. This depends, in part, on the typical size of  matrix elements of \Metpert. 

A first expectation\refs{\NVNLT} was that the typical size $\bar H^{\mu\nu}$ of the metric perturbation should be $\calo(1)$, in order to provide the needed rate \necrate\ for unitarization.  One way of thinking of this is that an $\calo(1)$ perturbation of the Hawking radiation is needed in order to transfer of order one qubit out of the BH per Hawking quantum.  

While this is possibly true, the preceding discussion has offered an attractive alternative.  Specifically, we have
\eqn\Hdef{H^{\mu\nu}(x)= \sum_\gamma O_\gamma f_\gamma^{\mu\nu}(x)\ ,}
where we recall that the $f_\gamma^{\mu\nu}$ are the  basis functions with $\calo(1)$ size in \Gexp, and $O_\gamma$ are given in \intops.  For nonviolence, these vary on scales $\Delta x\sim R$ (or, with a more general power of $R$).  
In a BH state $|\psi,T\rangle$, this gives expectation value
\eqn\Hexp{\langle\psi,T| H^{\mu\nu}|\psi,T\rangle = \sum_\gamma \langle\psi,T| O_\gamma |\psi,T\rangle f_\gamma^{\mu\nu}(x)\ .}
If $O_\gamma$ behave randomly with respect to the states $|\psi,T\rangle$, as discussed above, the normalizations described there then imply that the $O_\gamma$ have matrix elements $\sim 1/\sqrt{N}$, and thus that 
\eqn\Hsize{\langle\psi,T| H^{\mu\nu}(x)|\psi,T\rangle\sim {1\over \sqrt{N}}\ .}
One likewise sees that $H^n$ has size $\sim 1/\sqrt N$.
The nonviolence assumptions discussed in section three also indicate that these have time dependence on scales that grow with $R$, {\it e.g.} as $\sim R$.

Eq.~\Hsize\ thus suggests that the effective shift in the metric, in a ``typical" BH state, can be suppressed by a power of $1/\sqrt{N} = \exp\{-S_{bh}/2\}$, and thus is extremely tiny -- despite the information transfer constraint \necrate.  This possibility arises because the matrix elements of $H^{\mu\nu}$ are determined by the couplings $c_{A\gamma}$ of \Gexp\ to the individual BH states, which are very tiny, but the net effect of these couplings can be $\calo(1)$ because of the contribution of  $N=\exp\{S_{bh}\}$ states.  
A significant effect  arises because the BH has an enormous number of states that contribute to the new effects.  Indeed, this seems to be a generic way to enhance information transfer while maintaining small couplings, independent of the precise form of the couplings \Hinew\ and the conjecture \entrate.

One can also perturbatively estimate the effect of the couplings in $H_I$ of \HIexp\ on matter propagating near the BH, again using the formula \decrate, where now the initial and final states $|\alpha\rangle$ and $|\beta\rangle$ include matter scattering from the BH.  While the rate \decrate\ can be $\calo(1/R)$, that is for a change in momentum of the scattered matter that is  also $\calo(1/R)$.  Such a shift is negligible for matter accreting into a large black hole, or its radiation.\foot{However, gravitational radiation from a collision of black holes has typical momentum scales $\sim 1/R$, suggesting that such corrections could be significant in this context.  Exploration of this is left for future work.}  Preliminary analysis indicates that contributions that are higher order in $H_I$ are also significant, but since the energy/momentum transfers are effectively random, these are not expected to build up to a large effect.  It is interesting to contrast the case where the metric fluctuations behave ``classically," as in \NVNLT.  That can be described in the preceding formulas by taking $N\sim 1$.  Then the couplings $c_{A\gamma}$ are $\calo(1)$, and the metric fluctuation \Hsize\ is likewise $\calo(1)$.  In that case, higher order processes in $H_I$ are important, and can coherently build up many $\calo(1/R)$ energy/momentum transfers to produce, {\it e.g.}, an $\calo(1)$ deflection to trajectories of matter or light.

\subsec{Relation to previous work and ideas}

It is also illustrative to relate the preceding statements to previous work and common ideas on the subject.  Specifically, as noted, it is clear that a departure from the Hawking radiation state that is sufficient to lead to unitary evolution is in fact an $\calo(1)$ departure.  This was, for example, visible in Page's work \refs{\Pageone,\Pagetwo}, illustrating the large necessary departure in the von Neumann entropy; also, \refs{\Mathur} discussed this from the viewpoint of the detailed structure of the Hawking state.  It is worth emphasizing that the interactions described above are in accord with these statements; they in particular are of sufficient strength to produce an $\calo(1)$ correction to the state, as called for, for example, in Theorem 1 of \refs{\Mathur}.

However, it has also commonly been assumed that such $\calo(1)$ corrections to the state of the outgoing radiation require $\calo(1)$ corrections to the BH geometry.  The preceding argues that this is not the case.  Specifically, the interaction between the BH state and the atmosphere is an intrinsically quantum process, governed for example by the quantum hamiltonian $\HIexp$.  In such a process, one can get an $\calo(1)$ correction to the exterior state from  {\it small} interactions, due to the large number of BH states that can contribute to the total transition probability.  It bears emphasizing that, in the case of the universal coupling \Hiop, the quantity $H^{\mu\nu}$ correspondingly plays the role of a correction to the metric that depends on the BH quantum state, but that typical diagonal matrix elements of this operator are of a tiny size, \Hsize.  So, in effect, ``small" quantum corrections to the BH geometry are able to produce  the needed effect on the state.

\newsec{Future questions and observational tests}

If BHs obey the principles of quantum mechanics -- our first Postulate -- information must transfer out of a BH, no matter how large it is, or even more extreme modification of established physics is needed.  Either way, it appears that new effects are needed on scales $\sim R$ of the BH radius or larger.  These effects need to have $\calo(1)$ impact on the final state resulting from BH decay, in order to transform it from Hawking radiation with missing information $I\sim S_{BH}$ to a different state with no missing information.

\subsec{Possible observational probes of strong fluctuations}

It is interesting to seek observational tests for the presence of such new effects on scales $\sim R$, for example to distinguish between scenarios.  Our Postulates II-IV have led to a picture where there are metric fluctuations which extend outside the horizon some distance -- in the simplest picture a distance of size $R$.  {\it If} these fluctuations are ``strong," {\it i.e.} the typical metric fluctuation is of size $\Delta g_{\mu\nu}\sim 1$, as was na\"\i vely indicated by the need for an $\calo(1)$ effect, one would expect that they have a significant impact on motion near the BH, and could be searched for in observations sensitive to the near-horizon region.  
The exciting prospect of such observations\SGwind\ is becoming current reality, with our entry into the era of gravitational wave probes of near-horizon physics\refs{\LIGO,\SGLIGO}, {\it and} with the development of sensitivity to photon propagation near the horizon, with the Event Horizon Telescope (EHT)\refs{\EHT,\GiPs}.  

There is another question not yet answered by the Postulates, relevant to the possibility of observational tests. Specifically, while the arguments above lead to soft perturbations which may be strong, they do not, a priori, tell us at what point in a BH's evolution these perturbations become ``active." Previous arguments only lead to bounds.  For example, if information transfers out of a BH faster than a time scale $\sim R\log R$, that leads to contradictions in describing the experience of an observer who hovers outside, capturing information, and then falls into the BH\refs{\SuTh}.  On the other hand, information needs to begin to transfer out by a time of order $R^3$, if $S_{bh}\sim S_{BH}$, in order to have time to transfer the necessary entanglement\refs{\Pageone,\Pagetwo}.   For a BH such as Sgr A${}^*$, in the center of our galaxy, this range of times is $1\, hr$ to $10^{84}\,yr$.  For a solar mass BH, the range of times is $10^{-3}\,s$ to $10^{64}\,yr$.  

A more fundamental picture of the origin of the couplings \HiT, \eg\ possibly resulting from departures from the classical manifold/metric description of the BH spacetime, would, once we understood its dynamics, be expected to provide a prediction of this activation time scale.  Prior to having such a complete description, at least two possibilities are apparent. One is that the activation of the couplings \HiT\ could be a {\it saturational} effect, for example an effect that depends on the BH building up a large entanglement with its environment.  One might alternately describe this as the BH internal state  having most of its degrees of freedom excited from a nominal ground state.  If such saturation is responsible for producing significant couplings \HiT, the natural timescale to consider is $\sim R^3$, since this is the amount of time it takes to develop such a large entanglement through a collapse or Hawking process.  On the other hand, the effect responsible for the modification to GR could be {\it structural}:  the spacetime description fails to be accurate and complete once the strong gravitational region of the BH has formed.  If so, one would expect the relevant timescale to be comparable to the BH formation time, $\sim R\log R$.  

The latter, structural, time scale is certainly plausible, given questions about the quantum spacetime structure relevant to describing strong gravitational fields.  Moreover,  even for very large BHs like Sgr A${}^*$, their age, $\calo(10^9\, yr)$ is much longer than the lower bound $R\log R$.  These suggest it is reasonable to look for such effects; clearly their discovery would vindicate this approach, though a clear prediction for the activation time is needed for observation to strictly rule out the relevance of strong, soft metric perturbations.

Search for these effects via optical means is a cleaner approach than via gravitational waves, since the former simply requires description of propagation of light in a deformed background, whereas the latter, via LIGO, requires information about nonlinear evolution of the perturbations in order to make detailed predictions.  Searches for departures from GR templates in LIGO observations is very important in the latter context\refs{\SGLIGO,\SGobs}, but here we will focus on prospects for optical searches.

As \HiT\ and \Hiop\ show, the interactions may be described as fluctuations in the effective metric.  If the fluctuations have size $\calo(1)$ and scales $\sim R$, they will produce $\calo(1)$  deviations in geodesics.
This will in particular affect photons propagating near the BH.  This means that, with a candidate spectrum (\eg\ momenta, frequencies) for these fluctuations, one can examine the effects they would have on BH images, such as will be produced by EHT.  Specifically, the expected images are found by beginning with a model for the accreting matter and its radiation, and then using ray-tracing methods to follow the radiation outward and infer the image.  For  BH solutions without fluctuations, these methods produce images with distinct features, such as a BH shadow, and outside that a ring-like structure called the photon ring (see, {\it e.g.}, \refs{\PsJo} for discussion).  Strong, soft fluctuations, if present, are expected to add effectively random deflections to the photon trajectories, leading to the expectation of a smaller and fuzzier shadow, and distorted photon ring\refs{\SGfrank}.

An initial exploration of the modifications to these images due to strong, soft metric fluctuations was begun in the recent work \refs{\GiPs}.  With an example spectrum, with $\calo(1)$ fluctuations, this work demonstrated that the effect of these perturbations can be quite significant.  Taking as the typical time scale for the fluctuations $\tau\approx 8\pi^2 R$, one finds dramatic evolution of the image on this scale, as is shown in \GiPs\ and linked videos\refs{\Psvid}.  (Note that if instead $\tau$ is defined by the peak in the spectrum, it is three times smaller.) This is an important proof of principle for sensitivity, though visibility of such signals depends on the actual magnitude and time dependence of the fluctuations.  Note also that the time scale  $8\pi^2 R \sim 1$ hr for Sgr A${}^*$ is shorter than the ``averaging" scan time used by EHT, of order a few hours, while for its other primary target, M87, the time scale $8\pi^2 R$ is $\sim 60$ days.  This means that the latter appears to present greater prospects for directly investigating such time dependence\GiPs.

However, if no such effects are seen, the discussion of section 4.3 provides a possible explanation.

\subsec{Entropy-enhanced transfer}

Section four suggested  a scenario where unitarization -- an $\calo(1)$ effect -- is possible without having significant effect on matter propagating near the horizon; information transfer is enhanced, relative to the size of individual interactions, by the large entropy of the BH.  It is important to more carefully establish the viability of such a scenario.  

Specifically,  two important ingredients of this proposed scenario are 1) an information transfer rate governed by a formula like \entrate, such that information transfer of size \necrate\ is possible even with tiny couplings, \coupnorm, and 2) the statement that effects on matter near the BH depends on matrix elements like \Hsize, which are, due to the tiny couplings, highly suppressed.

An important element in the first ingredient is providing additional arguments or proof for a formula of the form  \entrate. Recall that the perturbative argument resulting in \decrate\ appears to provide significant support for this conjecture.

For the second ingredient, an important element is showing that the full evolution of the coupled subsystems -- BH and matter moving near the BH -- produces small perturbations in that matter.  The full evolution is determined, in the framework adopted in this paper, by a hamiltonian of the form \hamdiv, combining an internal hamiltonian $H_<$ which may have rather chaotic features, with an interaction hamiltonian $H_I$ of the form \Hinew\ or \HiT, and an exterior hamiltonian that is well-approximated as giving standard LQFT evolution.  Then, the important question is whether, with couplings of the size necessary for unitarization, the interactions $H_I$ have small effect on the outside matter.  Section four has argued that the typical perturbation in the effective metric is small, so is expected to have small effect, and a supplementary perturbative scattering argument was given.  It would nonetheless be nice to have a more complete analysis of the fully-coupled problem, proving the expected lack of enhancement in the full evolution.

It should be emphasized that, assuming that these effects are small, the basic mechanism responsible, relying on the large number of BH states, is not limited to metric couplings \HiT, but also can function via non-universal couplings \Hinew.

Given that the couplings necessary for unitarization may be small, another interesting question is whether they can arise from gravitational dressing corrections to the subsystem division that we described in section II, or whether they are a signal of truly new physics.  The former alternative potentially connects to the suggestion that soft hair plays a role in resolving  the puzzle\refs{\Hawksoft,\HPS}.  Indeed, a first expectation -- noted above -- is that gravitational dressing is too weak to make important corrections; a similar concern has led to skepticism about soft hair providing sufficiently large effects to effectively transfer information and restore unitarity.  But, the present discussion indicates the possibility that small couplings are sufficient, if they are of the right form, suggesting further exploration in these directions.

\subsec{Towards a fundamental picture}

The discussion of this paper has offered a very interesting new possibility for interactions that restore the reign of quantum mechanics over black holes, without having a large effect on matter propagating near a black hole.  Important tests of this scenario include both those of its logical consistency as well as observational tests checking whether matter near a black hole is significantly affected by new interactions with the black hole quantum atmosphere.  Beyond that, a more profound question is that of the underlying fundamental physics responsible for the deviations from quantum field theory that are necessary to save quantum mechanics.  As with the atom and the original development of quantum mechanics, models of correct black hole physics may provide a key guide to such more basic physics, which may well go beyond present knowledge of gravity.

So, this work presents a number of future directions.  One is sharpening the interplay between the constraints for necessary information transfer, the match to BH thermodynamics, the size of effects, and their observability.  Searching for these departures from GR predictions in EHT or LIGO observations is also clearly important.  And, finally, a fundamentally important question is to develop a more foundational picture of quantum spacetime, from which the interactions of this paper could emerge as an effective description.

\bigskip\bigskip\centerline{{\bf Acknowledgments}}\nobreak

I thank D. Berenstein, S. Britzen, W. Donnelly, J. Hartle, G. Horowitz, D. Marolf, D. Psaltis, and W. Van Dam for valuable discussions.  I also thank C. Nayak for valuable discussions, and for bringing \BHV\ to my attention.  I would like to thank the Albert Einstein Institute and the organizers of the conference {\sl Models of Gravity - Black Holes, Neutron Stars and the structure of space-time}, particularly S. Britzen, for an opportunity to present  preliminary accounts of this work.  
This work  was supported in part by the Department of Energy under Contract {DE}-{SC}0011702 and by  Foundational Questions Institute grant number FQXi-RFP-1507.

\appendix{A}{BH time slicings}

This appendix presents a brief discussion of time slicings that extend into the BH interior, such as are used in the main text.  Further discussion of dynamics on such slices is planned for future work\refs{\SGpr}.

First, note that  a $D$-dimensional nonrotating BH  can be described, including the region inside its future horizon, in Eddington-Finkelstein coordinates as
\eqn\EF{ds^2=-f(r) dv^2 + 2 dv dr + r^2 d\Omega^2_{D-2}\ .}
Here $f(r)=1-\mu(r)$ is a dimension-dependent function:  for two-dimensional BHs,
\eqn\tdbh{\mu(r)=e^{-2(r-R)}\ ,}
and for $D>3$
\eqn\hdbh{\mu(r)=\left({R\over r}\right)^{D-3}\ .}
In general, $\mu(r)=1$ at the horizon $r=R$, $\mu(r)$ vanishes at $r=\infty$, and $\mu(r)$ diverges at the singularity.

Time slices through these geometries can be defined by choosing a function $s(r)$, with $s(r)\rightarrow r$ as $r\rightarrow\infty$; then, for a given $T$, the corresponding slice is found as the solution of the equation
\eqn\slice{T=v-s(r)\ .}
At $r\rightarrow\infty$, these slices asymptote to slices of constant Schwarzschild time $t=T$.
Depending on the behavior of $s(r)$ for decreasing $r$, these slices can either intersect the singularity, or avoid it as with the nice slices of \refs{\LPSTU} (an explicit example is \refs{\BHQIUE}).  A particularly simple choice is $s(r)=r$, which we call ``straight" slices.  

In the coordinates $(T,r,\Omega)$, the metric takes the ADM form \ADMmet\ with
\eqn\slADM{N^2={1\over s'(2-fs')}\quad ,\quad N_r={1-fs'}\quad ,\quad q_{rr}= s'(2-fs')\ ,}
where $s'=ds/dr$.  Note that at the horizon $N^2=q_{ij}N^i N^j$.  In the straight slicing, 
\eqn\ssl{N^2={1\over 2-f}={1\over 1+\mu(r)}\quad ,\quad N_r={1-f}=\mu(r)\quad ,\quad q_{rr}= (2-f)=1+\mu(r)\ .}

In these slicings, the metric is independent of $T$, as was assumed in the main text.

\listrefs
\end